\documentclass[useAMS,usenatbib]{mn2e} 
\usepackage{graphicx}
\title{The effect of magnetic field on the inner Galactic rotation curve} 
\author[Chan \& Del Popolo]{Man Ho Chan$^1$ \thanks{chanmh@eduhk.hk}, Antonino Del Popolo$^{2,3}$
\\ $^1$Department of Science and Environmental Studies, The Education University of Hong Kong, Tai Po, Hong Kong \\
$^2$Dipartimento di Fisica e Astronomia, University of Catania, Viale Andrea Doria 6, 95125, Catania, Italy \\
$^3$Institute of Astronomy, Russian Academy of Sciences, Pyatnitskaya str. 48, 119017 Moscow, Russia}

\begin{document}

\date{Accepted XXXX, Received XXXX}

\pagerange{\pageref{firstpage}--\pageref{lastpage}} \pubyear{XXXX}

\maketitle

\label{firstpage}

\date{\today}

\begin{abstract}
In the past few decades, some studies pointed out that magnetic field might affect the rotation curves in galaxies. However, the impact is relatively small compared with the effects of dark matter and the baryonic components. In this letter, we revisit the impact of magnetic field on the rotation curve of our Galaxy. We show that the inner Galactic rotation curve could be affected significantly by the magnetic field. The addition of the inner bulge component, which has been proposed previously to account for the inner rotation curve data, is not necessary. The magnetic field contribution can fully account for the excess of the inner rotation velocity between 5 pc to 50 pc from the Galactic Centre. Our analysis can also constrain the azimuthal component of the central regular magnetic field strength to $B_0 \sim 50-60$ $\mu$G, which is consistent with the observed range.  
\end{abstract}


\section{Introduction}
The rotation curves of galaxies are important indicators of mass distribution in galaxies. Many rotation curves were revealed by the neutral atomic gas (e.g. HI, CO) which is believed to be a very good tracer of the gravitational field (e.g. Spitzer Photometry and Accurate Rotation Curves \citep{Lelli}). In particular, many past studies have shown that large-scale magnetic field can affect the gas dynamics in spiral galaxies \citep{Piddington,Nelson,Battaner,Battaner2}. Some studies even show that this magnetic field effect can explain the flatness of rotation curves in galaxies without the need of dark matter \citep{Nelson,Battaner,Battaner2}. This is known as the `magnetic alternative to dark matter' \citep{Sanchez}. Nevertheless, later studies have shown that the boost of rotation curves due to magnetic field contribution is less than 20 km/s at the outermost point of rotation curves \citep{Sanchez,Sanchez2}. Since then, the `magnetic alternative to dark matter' was no longer a popular model. 

In the past decade, the idea of the magnetic field effect on galactic rotation curves was revived. Some studies have shown that the magnetic field effect can explain why rotation curves in some galaxies start to rise again at the outer edges of the HI discs, such as our Galaxy \citep{Ruiz2} and the M31 galaxy \citep{Ruiz}. Considering the effects of magnetic field can somewhat improve the fits of the outer part of galactic rotation curves. However, some other studies have argued that the effect in the outer rotation curve region is not very significant \citep{Sanchez2,Elstner}.

Although the effect of magnetic field in the outer rotation curve region has been greatly debated, such effect in the inner region of a galaxy has not been discussed thoroughly. In this letter, we particularly investigate the magnetic field effect on the inner rotation curve of our Galaxy. There is a small rotation velocity excess range (between 5 pc to 50 pc from the Galactic Centre) which could not be accounted by the contributions of the supermassive black hole and the central bulge component \citep{Sofue}. An extra inner bulge has to be added to account for this abnormal excess range. We show that this small excess range could be explained by the magnetic field effect so that adding the extra inner bulge component is not necessary. 

\section{Magnetic field effect on rotation curve}
In a gaseous disc in equilibrium, magnetic field effects on the gas can be modelled as a pressure term in the asymmetric drift \citep{Sanchez2}. Such asymmetric drift is a consequence of the support by thermal, turbulent, cosmic ray and magnetic pressures. The dynamical effects of the regular magnetic field can significantly boost the gravitational orbital velocity due to the magnetic tension \citep{Nelson}. The total magnetic field in a galaxy can be simply expressed as a sum of the regular field term (the azimuthal component) $B_{\phi}$ and a random field term (the turbulent magnetic field component) $B_{\rm ran}$ \citep{Sanchez2,Elstner}. In particular, the random field can be isotropic or anisotropic. The contribution of the regular magnetic field component to the circular velocity is given by \citep{Ruiz}
\begin{equation}
v_{\rm B1}^2=\frac{r}{4\pi \rho_g} \left(\frac{B_{\phi}^2}{r}+\frac{1}{2} \frac{dB_{\phi}^2}{dr} \right),
\end{equation}
where $\rho_g$ is the gas density and $r$ is the radial distance from the Galactic Centre. The random magnetic field component would contribute to the circular velocity via the magnetic pressure term $P_B$ as \citep{Sanchez2}
\begin{equation}
v_{\rm B2}^2=\frac{r}{\rho_g}\frac{dP_B}{dr}=\frac{r}{\rho_g}\frac{d}{dr} \frac{\langle B_{\rm ran}^2 \rangle}{8 \pi},
\end{equation}
where $\langle B_{\rm ran}^2 \rangle$ is the mean-square value of the random magnetic field strength. Therefore, the total contribution of the magnetic field to the circular velocity is:
\begin{equation}
v_{\rm mag}^2=v_{\rm B1}^2+v_{\rm B2}^2=\frac{r}{8\pi \rho_g} \left[\frac{2B_{\phi}^2}{r}+\frac{d}{dr}(B_{\phi}^2+\langle B_{\rm ran}^2 \rangle) \right].
\end{equation} 

In the inner Galactic Centre region ($r \le 500$ pc), the rotation curve is also contributed by the supermassive black hole $v_{\rm BH}$ and the baryonic bulge $v_{\rm bulge}$ components. Therefore, including the magnetic field contribution, the observed total rotation curve is
\begin{equation}
v^2=v_{\rm BH}^2+v_{\rm bulge}^2+v_{\rm mag}^2.
\end{equation}
The supermassive black hole rotation curve contribution is 
\begin{equation}
v_{\rm BH}^2=\frac{GM_{\rm BH}}{r},
\end{equation}
where $M_{\rm BH}=(4.154 \pm 0.014) \times 10^6M_{\odot}$ is the mass of the supermassive black hole \citep{Abuter}. The bulge mass density can be modeled by the exponential spheroid model with scale radius $a$ as \citep{Sofue}:
\begin{equation}
\rho_{\rm bulge}(r)=\rho_ce^{-r/a},
\end{equation}
where $\rho_c$ is the central bulge mass density. Therefore, the bulge rotation curve contribution is given by
\begin{equation}
v_{\rm bulge}^2=\frac{GM_0}{r}F \left(\frac{r}{a}\right),
\end{equation}
where $M_0=8\pi a^3\rho_c$ and $F(x)=1-e^{-x}(1+x+x^2/2)$ \citep{Sofue}. We take the values $M_0=8.4\times 10^9M_{\odot}$ and $a=0.12$ kpc obtained in \citet{Sofue} to perform our analysis.

To match the inner Galactic rotation curve data without considering the magnetic field contribution, two bulge components (inner bulge and outer bulge) were assumed in previous studies \citep{Sofue}. However, in the followings, we will investigate whether the magnetic field contribution can mimic the effect of the inner bulge component. Therefore, we assume that there is only one bulge component to minimise the number of parameters in our analysis.

Moreover, the dark matter contribution is not considered here because it is not significant in the inner Galactic Centre region ($r \le 500$ pc). Assuming the Navarro-Frenk-White (NFW) profile \citep{Navarro}, the total mass of dark matter is less than 5\% of the bulge mass inside 500 pc, with the fitted parameters in \citet{Sofue2}. Therefore, we neglect the contribution of dark matter for simplicity. We also neglect the gravitational contribution of gas in the inner Galactic Centre region as the gas mass is less than 3\% of the bulge mass inside 500 pc \citep{Ferriere}.

Over $\sim 300$ pc along the Galactic plane and $\sim 150$ pc in the vertical direction at the Galactic Centre, the magnetic field is approximately horizontal and the strength is very strong, which can range from $B \sim 0.01-1$ mG for general intercloud medium \citep{Ferriere2}. The equipartition between magnetic field energy and turbulent energy suggests that $B \propto \rho_g^{1/2}$ \citep{Schleicher}. This is also supported by the observed relation between magnetic field and star-formation rate \citep{Heesen,Tabatabaei}. Therefore, we assume that the regular magnetic field profile follows the exponential spheroid model as
\begin{equation}
B_{\phi}=B_0e^{-r/2a},
\end{equation}
where $B_0$ is the central regular magnetic field strength. Note that Eq.~(8) follows from Eq.~(6) only if gas density is assumed to be proportional to the baryonic bulge mass density $\rho_{\rm bulge}(r)$. For the random magnetic field, it is an important component in the interstellar medium of galaxies \citep{Beck}. We define $\eta$ as the ratio of the regular magnetic field to the total magnetic field so that the random magnetic field strength can be expressed as $\langle B_{\rm ran}^2 \rangle=(\eta^{-2}-1)B_{\phi}^2$. Although some earlier studies found that the ratio is around $\eta \sim 0.6-0.7$ in some galaxies \citep{Fletcher,Beck2,Sanchez2}, recent studies have shown that these ordered fields are dominated by anisotropic random fields and the ratio should be $\eta \sim 0.01-0.3$ in galactic disk region \citep{Beck}. Nevertheless, the actual value of $\eta$ at the Galactic Centre is uncertain and the value of $\eta$ may be larger. In the followings, we will first assume $\eta=0.65$. Then, we will also demonstrate the cases for $\eta=0.3$ and $\eta=0.9$ for comparison.   

For the gas density, observational data show that the gas number density is close to $n_H \sim 10-100$ cm$^{-3}$ at the inner Galactic Centre and $n_H \sim 1$ cm$^{-3}$ out to $\sim 220$ pc along the Galactic plane \citep{Ferriere,Ferriere2}. We follow \citet{Ferriere} to assume that the gas density decreases exponentially with $r$ for small $r$ and then approaches a constant value $\rho_0'$ when $r$ becomes large. Therefore, we write the gas density profile as
\begin{equation}
\rho_g=\rho_0 \exp \left(-\frac{r}{1~\rm pc} \right)+\rho_0'.
\end{equation}

Putting Eq.~(8) and Eq.~(9) into Eq.~(3), we get
\begin{equation}
v_{\rm mag}^2=\frac{v_0^2e^{-r/a}[1-\eta^{-2}(r/2a)]}{\exp(-r/1~{\rm pc})+y},
\end{equation}
where $v_0^2=B_0^2/(4 \pi \rho_0)$, $y=\rho_0'/\rho_0$. Here, $v_0$ and $y$ are free parameters in our model.

\section{Data analysis}
The inner rotation curve data have been obtained in \citet{Sofue}. We will focus on the region $r \le 360$ pc because the rotation curve attains its maximum at around $r=360$ pc. At this position, the percentage contribution of the bulge component is maximised. Larger than $r=360$ pc, the dark matter and disc components start to contribute more to the Galactic rotation curve. We fit our predicted total rotation curve $v(r)$ with the observed data $v_{\rm obs}(r)$. To quantify the goodness of fits, we calculate the reduced $\chi^2$ value of the fits, which is defined as
\begin{equation}
\chi_{\rm red}^2=\frac{1}{N-M}\sum_{i=1}^{N} \frac{(v_i-v_{{\rm obs},i})^2}{\sigma_i^2},
\end{equation}
where $\sigma_i$ is the uncertainty of the observed rotation curve data, $N$ is the total number of data points and $M$ is the number of free parameters. Here, we have $M=2$. 

In Fig.~1, we present the best fit of our model (with $\eta=0.65$) and separate the corresponding rotation curve contributions. The best-fit values are $v_0=16.4$ km/s and $y=0.023$, with $\chi_{\rm red}^2=0.14$. We can see that including the magnetic field contribution could provide an excellent fit to the observed inner rotation curve without introducing any inner bulge component suggested in \citet{Sofue}. The reduced $\chi^2$ value for adding an inner bulge component without magnetic field contribution is $\chi_{\rm red}^2=0.12$, almost the same goodness of fit. Note that the contribution of the magnetic field effect to the rotation velocity could be slightly negative when $r>2\eta^2a \approx 0.1$ kpc. Fig.~2 shows the comparison of the residual plots between the Galactic total rotation curve data and the two scenarios. No systematic trend of the residuals is shown for both scenarios.

At the Galactic Centre, the gas density is $\rho_g \approx 1.4m_pn_H$, where $m_p$ is the proton mass. If we take the asymptotic number density $n_H=1$ cm$^{-3}$ \citep{Ferriere}, we have $\rho_0' \approx 2.3 \times 10^{-24}$ g/cm$^3$. Using the best-fit values $y=0.023$ and $v_0=16.4$ km/s, we get $\rho_0=1.0\times 10^{-22}$ g/cm$^{-3}$ (i.e. $n_H \approx 43$ cm$^{-3}$) and $B_0=58$ $\mu$G. These values are consistent with the number density observed and the central total magnetic field constrained ($B \sim 10-100$ $\mu$G) \citep{Ferriere,Ferriere2,Guenduez}.

As the actual value of $\eta$ at the Galactic Centre is uncertain, we also investigate the cases for $\eta=0.3$ and $\eta=0.9$. For $\eta=0.9$, we also get a very good fit with $v_0=15.2$ km/s ($B_0=54$ $\mu$G) and $y=0.023$ ($\chi_{\rm red}^2=0.11$). However, for $\eta=0.3$, a relatively poor fit is obtained ($\chi_{\rm red}^2=1.46$). We plot the corresponding components and the total rotation curves fitted in Fig.~3. Generally speaking, for $\eta>0.6$, a very good fit would be obtained ($\chi_{\rm red}^2<0.17$).

\begin{figure}
\vskip 10mm
 \includegraphics[width=80mm]{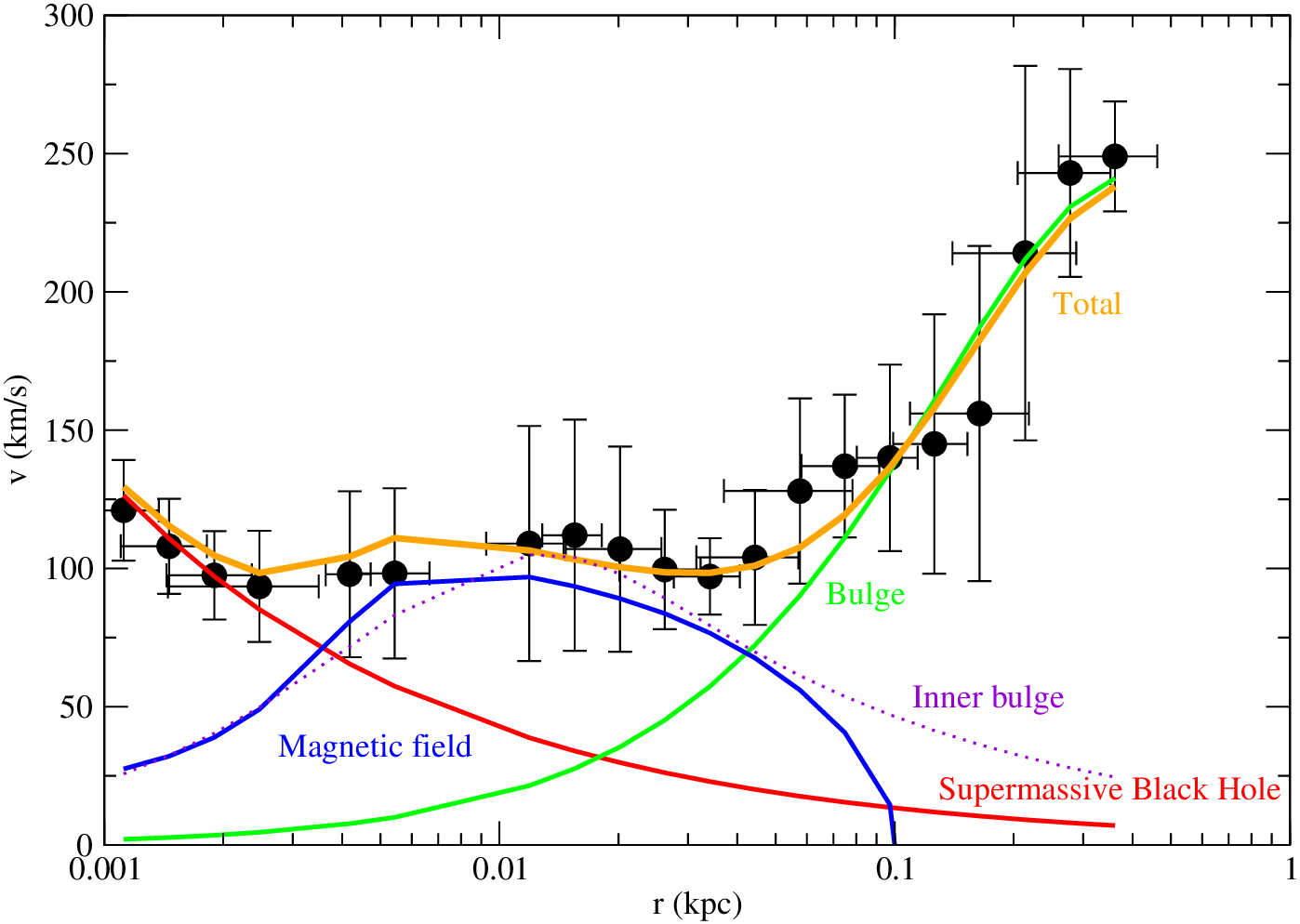}
 \caption{The black dots with error bars represent the inner Galactic rotation curve data for $r=1-360$ pc \citep{Sofue}. The red, green, and blue solid lines indicate the best-fit rotation curve components of the supermassive black hole, bulge, and magnetic field contribution respectively. The orange line is the best-fit total rotation curve in our model. The violet dotted line represents the inner bulge contribution assumed in \citet{Sofue}. Here, we have assumed $\eta=0.65$.}
\vskip 10mm
\end{figure}

\begin{figure}
\vskip 10mm
 \includegraphics[width=80mm]{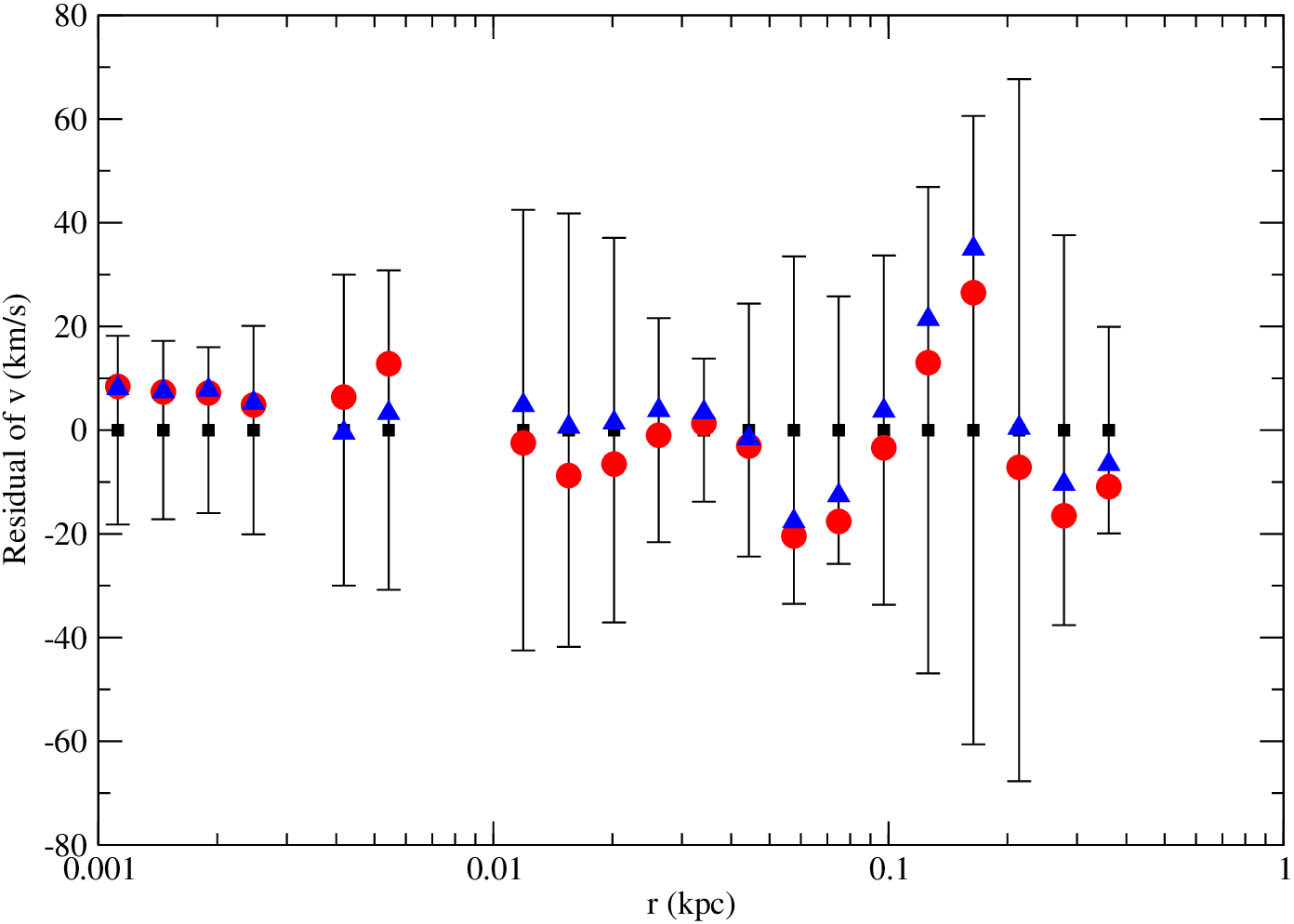}
 \caption{The red circles and blue triangles represent the residuals of the best-fit total rotation curve due to the contributions of the magnetic field effect and the inner bulge component respectively compared with the inner Galactic rotation curve data.}
\vskip 10mm
\end{figure}

\begin{figure}
\vskip 10mm
 \includegraphics[width=80mm]{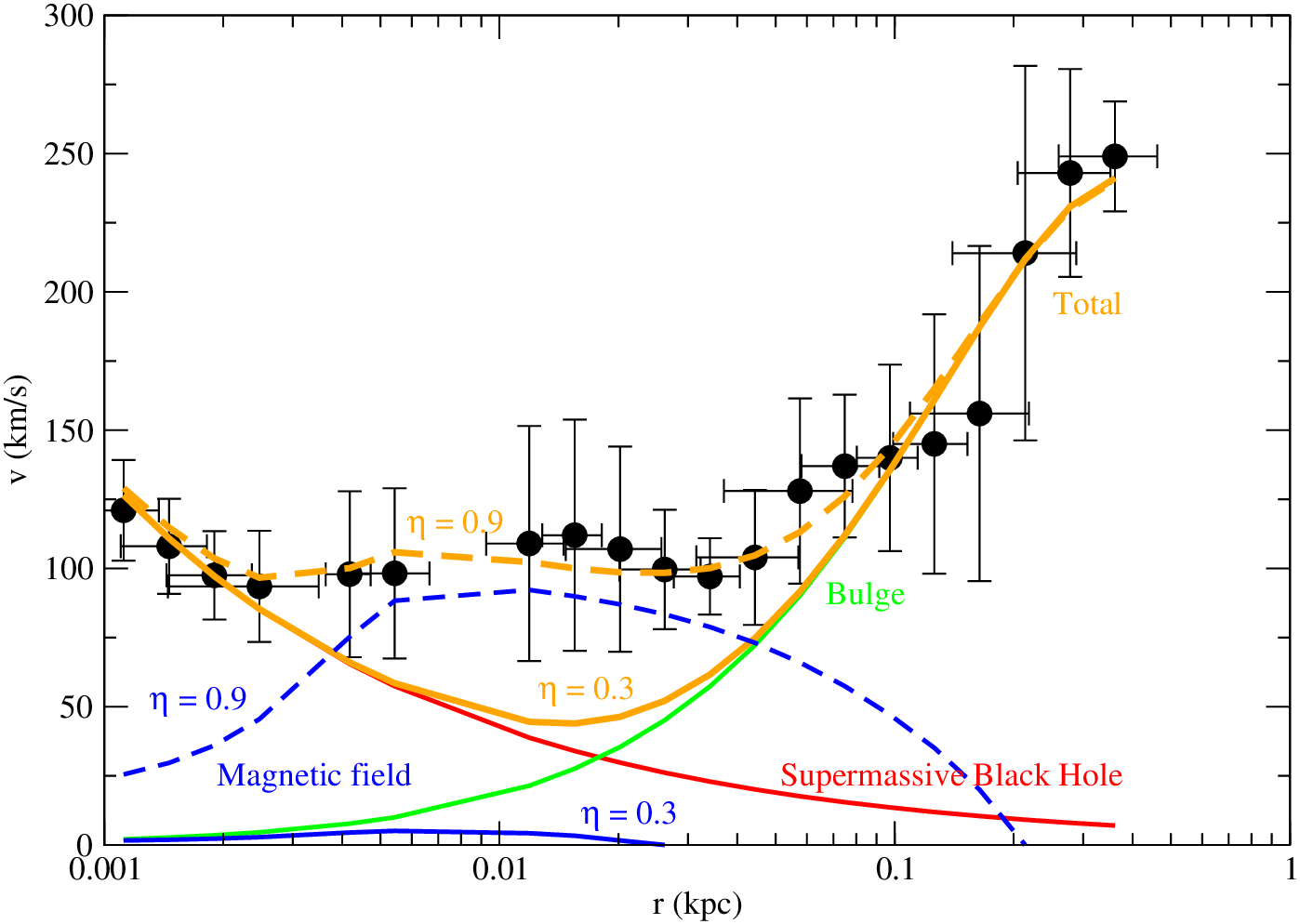}
 \caption{The black dots with error bars represent the inner Galactic rotation curve data for $r=1-360$ pc \citep{Sofue}. The red, green, and blue lines indicate the best-fit rotation curve components of the supermassive black hole, bulge, and magnetic field contribution respectively (blue solid: $\eta=0.3$; blue dashed: $\eta=0.9$). The orange lines are the best-fit total rotation curve in our model (orange solid: $\eta=0.3$; orange dashed: $\eta=0.9$).}
\vskip 10mm
\end{figure}

\section{Discussion}
In this letter, we show that adding the magnetic field contribution can satisfactorily explain the inner Galactic rotation curve data without invoking an inner bulge component. The magnetic field effect on rotation curve is a predicted effect in magneto-hydrodynamics (MHD). Our results provide an indirect evidence that magnetic field can affect the galactic rotation curve significantly. Previous studies have shown that the magnetic field contribution cannot boost the gas rotating speed by more than 20 km/s in the outermost region \citep{Sanchez2}. Now we show that such effect can be large in the central region of a galaxy. The maximum contribution of the magnetic field on the rotation curve is 97 km/s at $r \sim 12$ pc. 

In our analysis, we have assumed that magnetic field strength traces the baryonic distribution (i.e. the bulge density), which is predicted by theoretical models. For example, numerical simulations and the equipartition theory show that the magnetic field strength follows the baryonic density in galaxy clusters and galaxies \citep{Dolag,Govoni,Schleicher} and it is supported by observational data \citep{Heesen,Tabatabaei,Weeren}. Our constrained magnetic field strength at the Galactic Centre ($B_0 \sim 50-60$ $\mu$G) is also consistent with the order of magnitude of the observed total magnetic field strength $B \sim 10-100$ $\mu$G \citep{Ferriere2,Guenduez}. These could be verified by future observational data of magnetic field at the Galactic Centre. Moreover, we have first taken the value of $\eta$ to be a constant $\eta=0.65$. Some other studies have found that the magnetic field strength might be dominated by anisotropic random field rather than the regular field \citep{Houde,Beck}. The value of $\eta$ can be as small as $\sim 0.1$ for the disk region in spiral galaxies \citep{Beck}. However, some studies have revealed a very large regular field strength $\sim 1$ mG at the Galactic Centre \citep{Eatough}. Therefore, the actual value of $\eta$ in the Galactic Centre region is uncertain. We have particularly investigated the cases of $\eta=0.3$ and $\eta=0.9$ for comparison. We have found that $\eta=0.9$ can also give a good fit for the data. Generally, $\eta>0.6$ could provide good fits with the rotation curve data without invoking the inner bulge component. Further radio observations are definitely required to examine the value of $\eta$ as well as our model presented. 

On the other hand, we have also assumed that the gas density follows the exponential density profile in the deep central region and approaches a constant value at a relatively large $r \sim 100$ pc. The constant gas density at $r \sim 100$ pc is supported by observational data \citep{Ferriere,Ferriere2}. We have also tried the isothermal density profile, which is commonly used as a model for interstellar medium \citep{Kalashnikov}, to model the gas density distribution. A good fit can still be obtained (with $B_0=62$ $\mu$G and $\chi_{\rm red}^2=0.12$). Therefore, our results are not very sensitive to the gas density profile assumed.

To conclude, we show that magnetic field can affect inner rotation curve significantly. We anticipate that such effect can also be seen in the inner region of other galaxies. Future high-quality observations of the inner galactic rotation curves could verify our suggestion.

\section{acknowledgements}
We thank the anonymous referee for useful constructive feedback and comments. This work was partially supported by the Seed Funding Grant (RG 68/2020-2021R) and the Dean's Research Fund (activity code: 04628) from The Education University of Hong Kong.

\section{Data availability statement}
The data underlying this article will be shared on reasonable request to the corresponding author.

\label{lastpage}

\end{document}